\documentclass[prl,twocolumn,superscriptaddress,floatfix]{revtex4-1}
\usepackage{amsmath,amsfonts,amssymb,amsthm,graphics,graphicx,epsfig,times}
\usepackage[colorlinks=true,citecolor=blue,linkcolor=blue]{hyperref}
\usepackage[usenames]{color}
\usepackage{subfigure}
\usepackage{dcolumn}
\usepackage{bm}
\usepackage{color}
\usepackage{verbatim}
\usepackage{epstopdf}
\usepackage{amstext}
\usepackage{latexsym}
\usepackage{hyperref}
\usepackage{amsfonts}
\usepackage{psfrag}
\usepackage{xcolor}

\newcommand{\ket}[1]{\ensuremath{\left|{#1}\right\rangle}}
\newcommand{\bra}[1]{\ensuremath{\left\langle{#1}\right|}}

\usepackage{mathptmx} 

\DeclareMathAlphabet\mathcal{OMS}{cmsy}{m}{n}

\begin{document}

\title{Quantum Rabi model in the Brillouin zone with ultracold atoms}

\author{Simone Felicetti}
\affiliation{Laboratoire Mat\' eriaux et Ph\' enom\`enes Quantiques, Sorbonne Paris Cit\' e, Universit\' e Paris Diderot, CNRS UMR 7162, 75013, Paris, France}
\author{Enrique Rico}
\affiliation{Department of Physical Chemistry, University of the Basque Country UPV/EHU, Apartado 644, E-48080 Bilbao, Spain}
\affiliation{IKERBASQUE, Basque Foundation for Science, Maria Diaz de Haro 3, E-48013 Bilbao, Spain}
\author{Carlos Sabin}
\affiliation{Instituto de F\'isica Fundamental, CSIC, Serrano 113-bis, E-28006 Madrid, Spain}
\author{Till Ockenfels}
\affiliation{Institut f\"ur Angewandte Physik der Universit\"at Bonn, Wegelerstr. 8, D-53115 Bonn, Germany}
\author{Johannes Koch}
\affiliation{Institut f\"ur Angewandte Physik der Universit\"at Bonn, Wegelerstr. 8, D-53115 Bonn, Germany}
\author{Martin Leder}
\affiliation{Institut f\"ur Angewandte Physik der Universit\"at Bonn, Wegelerstr. 8, D-53115 Bonn, Germany}
\author{Christopher Grossert}
\affiliation{Institut f\"ur Angewandte Physik der Universit\"at Bonn, Wegelerstr. 8, D-53115 Bonn, Germany}
\author{Martin Weitz}
\affiliation{Institut f\"ur Angewandte Physik der Universit\"at Bonn, Wegelerstr. 8, D-53115 Bonn, Germany}
\author{Enrique Solano}
\affiliation{Department of Physical Chemistry, University of the Basque Country UPV/EHU, Apartado 644, E-48080 Bilbao, Spain}
\affiliation{IKERBASQUE, Basque Foundation for Science, Maria Diaz de Haro 3, E-48013 Bilbao, Spain}

\begin{abstract}
The quantum Rabi model describes the interaction between a two-level quantum system and a single bosonic mode. We propose a method to perform a quantum simulation of the quantum Rabi model introducing  a novel implementation of the two-level system, provided by the occupation of Bloch bands in the first Brillouin zone by ultracold atoms in tailored optical lattices. The effective qubit interacts with a quantum harmonic oscillator implemented in an optical dipole trap. Our realistic proposal  allows to experimentally investigate the quantum Rabi model for extreme parameter regimes, which are not achievable with natural light-matter interactions. Furthermore, we also identify a generalized version of the quantum Rabi model in a periodic phase space.
\end{abstract}

\date{\today}

\maketitle

The Rabi model~\cite{Rabi36} is a semiclassical description of the dipolar interaction of a nuclear spin with electromagnetic radiation. Its full quantum version, known as the quantum Rabi model, has been applied more generally to describe the interaction between a two-level quantum system and a single bosonic mode, regardless of their specific physical origin. In the strong coupling  regime, where the coupling strength is larger than dissipation rates but small compared to the system characteristic frequencies, the quantum Rabi model can be reduced via a rotating-wave approximation to the Jaynes-Cummings model~\cite{Jaynes63}. The latter has been used for decades to explain a plethora of experiments~\cite{ParisCQEDbook13,InnsbruckReview08,Wallraff04} in quantum optics and condensed matter, such as cavity quantum electrodynamics (QED), trapped ions, circuit QED and quantum dots. 

More recently, it has been experimentally demonstrated that the ultrastrong coupling regime can also be achieved~\cite{Bourassa2009,Niemczyk2010,Fedorov2010,Diaz2010,Anappara2009,Gunter2009}, where the coupling strength is large enough to break the rotating-wave approximation and the full quantum Rabi model must be considered. The interest in the ultrastrong coupling regime is motivated by novel fundamental features~\cite{Ciuti2005,Ciuti2006,Ridolfo2012,Felicetti2014,Stassi2015} and potential computational benefits~\cite{Rossatto2016,Nataf2011,Romero2012,Kyaw2014}. Despite its ubiquity, analytical solutions for the quantum Rabi model spectrum has been developed only recently~\cite{Braak2011}, prompting further theoretical efforts to study generalizations of the quantum Rabi model, including anisotropic couplings~\cite{Xie2014}, two-photon interactions~\cite{Travenec2012,Albert2011,Felicetti2015} and the Dicke model~\cite{Braak2013}. Besides, significant efforts are devoted to reproducing these models using different quantum technologies~\cite{Baumann2010,Ballester2012,Mezzacapo2014,Pedernales2015,Klinder2015}.

Ultracold atoms represent one of the most advanced quantum platforms for the implementation of analog quantum simulations~\cite{Bloch2008}. They have mostly been associated with the implementation of quantum many-body and condensed-matter models.  Spin-like degrees of freedom have been implemented with ultracold atoms using internal electronic transitions~\cite{Demler2002,Aikawa2014,Lucke2014}. Alternatively,  the creation of two-level systems with atomic quantum dots~\cite{recati2005atomic, orth2008dissipative} or double-well potentials~\cite{cirone2009collective} have been proposed. Remarkably, relativistic effects~\cite{witthaut2011effective,salger2011klein,Gonzalez2014} have also been simulated using ultracold atoms.

Here, we propose a quantum simulation of the quantum Rabi model with cold atoms loaded onto a periodic lattice.  The effective two-level quantum system is simulated by different Bloch bands in the first Brillouin zone, and the bosonic mode is represented by the motion of the atomic cloud in a superimposed harmonic optical-trap potential.  The qubit energy spacing is proportional to the periodic lattice depth, while the interacting bosonic mode is intrinsic in the qubit definition.  When the edge of the Brillouin zone is reached, we find that a generalized version of the quantum Rabi model in periodic phase space is realized.

\begin{figure}[]
\centering
\includegraphics[angle=0, width= 0.4\textwidth]{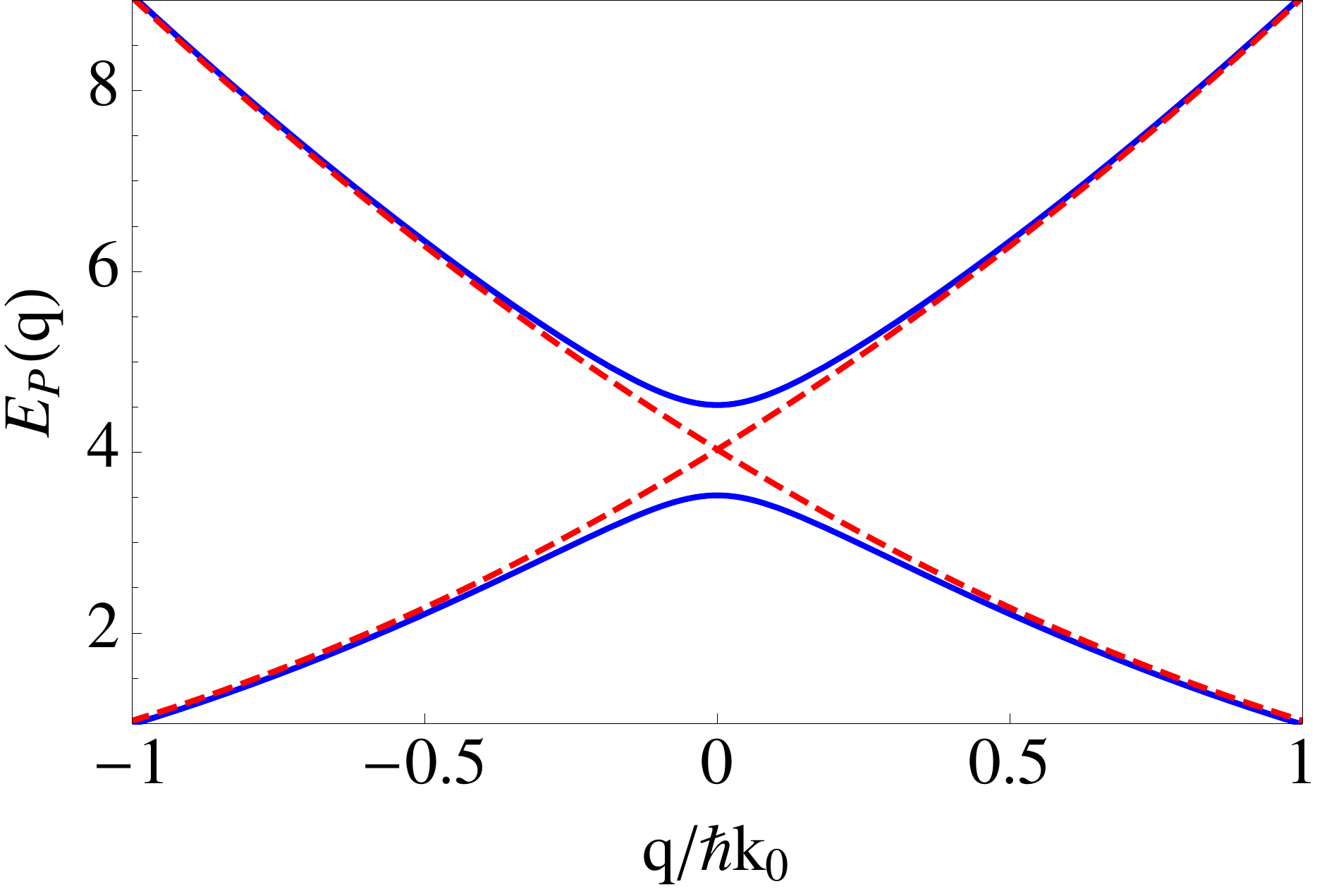}
\caption{\label{bandstructure} Band structure for an optical lattice potential. Comparison of the dispersion relation for the first and second bands for a lattice-potential depth of $V= 2 E_r$ (solid blue line) and a free particle (dashed red line). The gap between the first and second bands corresponds to the effective qubit energy splitting.}
\end{figure}

We show that our method, feasible with nowadays technology, can access extreme parameters regimes of the quantum Rabi model. This will allow for the experimental study of the transition between the deep strong coupling (DSC) regime \cite{Casanova2010}, where the coupling strength is larger than bosonic-mode frequency, and the dispersive deep strong coupling (dDSC) regime, where in addition the frequency of the qubit is much larger than the frequency of the bosonic mode. The complexity of such a model has been recently highlighted by the prediction of a phase transition, even in the single-qubit case, when the ratio between the qubit and the bosonic frequency tends to infinity~\cite{Hwang2015}.

The system here considered is composed of a cloud of ultracold atoms exposed to two laser-induced potentials:  a periodic lattice and a harmonic trap. When the atom density is sufficiently low, interactions among the atoms are negligible, and the system can be described with a  single-particle Hamiltonian, composed of the sum of a harmonic part $H_{\rm P}$ and a quadratic term
\begin{equation}
\hat H =  \hat H_{\rm P} + \frac{m \omega^{2}_0}{2} \hat x^2\ , \quad \quad  \hat H_{\rm  P} =  \frac{ \hat p^2}{2m}  + \frac{V}{2}\cos{ (4 k_0 \hat x) }, 
\label{fullHam}
\end{equation}
where, $ \hat p= -i\hbar \frac{\partial}{\partial x}$ and $\hat x$ are momentum and position of an atom of mass $m$, respectively. Here, $\omega_0$ is the angular frequency of the atom motion in the harmonic trap, while $V$ and $4k_0$ are the depth and wave-vector of the periodic potential, respectively. The periodic lattice is resulting from a four-photon interaction with a driving field~\cite{ritt2006fourier, witthaut2011effective, salger2009directed} of wave-vector $k_0$. 

In the following, we will assume that the harmonic trap is slowly varying on the length-scale of the periodic potential. Under this assumption, the most suitable basis is given by the Bloch functions $\langle x \ket{\phi_{n}(q)} = \phi_n(q,x) = e^{iq x/\hbar} u_{n_b}(x)$, with $q$ the quasi-momentum and $n_b$ is the band index, while $u_{n_b}(x)$ must be a periodic function with the same periodicity of the periodic potential. Accordingly, we define $u_{n_b}(x) = e^{-i2k_0x} e^{i 4n_b k_{0} x}$ where we have added the phase $e^{-i2k_0x} $ to the $u_{n_b}(x)$ functions definition in order to obtain a convenient first Brillouin zone, $q\in ( -2\hbar k_0, 2\hbar k_0 ] $. Notice that the Bloch functions are identified by a discrete quantum number, the band index  $n_b$, and a continuous variable, the quasi-momentum $q$. Hence, we can define a continuous and a discrete degrees of freedom, and rewrite the Bloch basis as $\ket{\phi_{n_b}(q)} = \ket{q}\ket{n_b}$.

First, let us consider the periodic part $\hat H_p$ of the system Hamiltonian, later we will discuss the effect of the harmonic trap. It is straightforward to see that the momentum operator is diagonal in the Bloch basis, while the periodic potential introduces a coupling between adjacent bands
\begin{eqnarray}
\label{bands}
\hat H_p \ket{q}\ket{n_b} &=& \frac{1}{2m}\big[ q + (2n_b -1 )2\hbar k_0 \big]^2 \ket{q}\ket{n_b} \\ \nonumber
&\ & + \frac{V}{4}\bigg( \ket{q}\ket{n_b +1}  +\ket{q}\ket{n_b-1}  \bigg) .
\end{eqnarray}

Let us now include the quadratic term of Eq.~\eqref{fullHam} in our treatment. In the Bloch basis, we can write
\begin{equation}
\langle \tilde{q}, \tilde{n}_{b} | \hat x^2 |q,n_b \rangle =  \int^{+ \infty}_{-\infty} dx \, x^{2} e^{i \left[ 4 (n_b - \tilde{n}_{b} )k_{0} + (q-\tilde{q}) /\hbar \right] x}.
\label{xsquareint}
\end{equation}
Considering diagonal elements in the band index, i.e., setting $\tilde{n}_{b} = n_b$, we have $\langle \tilde{q}, n_b | \hat x^2 |q,n_b \rangle = - \hbar^{2} \langle \tilde{q}, n_b |  \frac{\partial^{2}}{\partial q^{2}} |q,n_b \rangle$. Hence, we see that the harmonic potential introduces an operator, diagonal in the qubit Hilbert space, which can be expressed as $\hat x = -i\hbar \frac{\partial}{\partial q}$, in the Bloch basis. This allows us to define the quasi-momentum operator $\hat q$ and the position operator $\hat x$, which satisfy the commutation relation $\left[ \hat x, \hat q\right] = i\hbar$.

On the other hand, for $\tilde{n}_{b} \neq n_b$, the integral in Eq.~\eqref{xsquareint} is maximized if the relation $4 \hbar k_{0} (n_b - \tilde{n}_{b} ) = \tilde{q} - q$ is satisfied. Hence, the quadratic potential introduces a coupling between neighboring bands, for states whose momenta satisfy $\tilde{q} - q = 4 \hbar k_{0}$, of the kind $\left( \ket{2\hbar k_0, n_{b}}\bra{-2\hbar k_0, n_{b}+1} + {\text{H.c.}} \right)$. This effective coupling is due to the periodicity of the quasi-momentum, which mixes the bands at the boundaries of the Brillouin zone. Such a coupling can be neglected as far as the system dynamics involves only values of the quasi-momentum $\hat q$ included within the first Brillouin zone.

Assuming that the system dynamics is restricted to the two bands with lowest energy ($n_b=0,1$), as shown in Fig.~\ref{bandstructure},
the periodic part $\hat H_{\rm P}$ of the Hamiltonian can be rewritten in the Bloch basis as
\begin{equation}
\label{implementrabi}
\hat H_{\rm P}(q) = \frac{1}{2m}\left( \begin{matrix} q^2 + 4\hbar k_0\  q & 0  \\ 0 & q^2 - 4\hbar k_0\  q \end{matrix}\right) + \frac{V}{4} \left( \begin{matrix}   0 & 1  \\ 1 & 0 \end{matrix}\right),
\end{equation}
while the quadratic trapping potential
\begin{equation}
\hat x^2 = -\hbar^{2} \frac{\partial^{2}}{\partial q^{2}} \begin{pmatrix} 1 & 0  \\ 0 & 1 \end{pmatrix} + \Omega \left( \ket{2\hbar k_0, 0}\bra{-2\hbar k_0, 1} + {\text{H.c.}} \right),
\end{equation}
where $\Omega = \langle -2\hbar k_0, 1 |\hat{x}^{2} |2\hbar k_0, 0 \rangle$. Defining annihilation operator $\hat a = \sqrt{\frac{m \omega_0}{2\hbar}} \left( -i\hbar \frac{\partial}{\partial q} + \frac{i}{m\omega_0} \hat q \right) $ and creation operator $\hat a^\dagger$, respectively, and rotating the qubit basis with the unitary operator
$U = \frac{1}{\sqrt{ 2}}
\left( \begin{matrix}
1 & -1 \\ 1 & \ \ \ 1
\end{matrix} \right)$, the total system Hamiltonian, up to the umklapp term, can be finally rewritten as
\begin{equation}
\hat H = \hbar \omega_0 \hat a^\dagger \hat a + \frac{\hbar \omega_q}{2}\sigma_z +  i \hbar g \sigma_x \left( a^\dagger -  a  \right),
\label{standardRabi}
\end{equation}
which corresponds to the quantum Rabi Hamiltonian where we have defined the effective qubit energy spacing $\omega_q = \frac{V}{2\hbar}$ and the interaction strength $g = 2 k_0 \sqrt{\frac{\hbar \omega_0}{2m}}$. The Pauli matrices are defined in the rotated basis and, using the notation for the Bloch bands, they can be written as
\begin{eqnarray}
\label{pauliformalism}
\sigma_x \  &=&\ \ket{n_b=0}\bra{n_b=0} - \ket{n_b=1}\bra{n_b=1}, \\ \nonumber
\sigma_z \  &=&\ \ket{n_b=1}\bra{n_b=0} + \ket{n_b=0}\bra{n_b=1}. 
\end{eqnarray}
Notice that, in the standard form of the quantum Rabi model, the qubit-field coupling is usually written in terms of the position operator, while in Eq.~\eqref{standardRabi} it appears in terms of the momentum operator. The two definitions are equivalent up to a global phase factor.

The full system Hamiltonian of Eq.~\eqref{fullHam} resembles the quantum Rabi model only when the effective coupling between different bands induced by the harmonic potential can be neglected. Such an approximation holds as long as
the system wave-function $\langle q \ket{\psi(t)}$ is completely included in the first Brillouin zone.
Clearly, this constraint limits the proposed implementation to values of the momentum $\hat q$ smaller in modulus than $2\hbar k_0$.
In the following, we will show that  this constraint does not impede to observe the highly non-trivial behavior of the quantum Rabi model in the DSC and dDSC regimes. The DSC regime was introduced theoretically in \cite{Casanova2010} and it is generically characterized by $g>\omega_0$. However, in the hitherto unexplored dDSC regime, we have the condition $\omega_q \gg\omega_0$. Interestingly, some key features of the DSC regime are reproduced even when the periodicity of the quasi-momentum becomes relevant for the system dynamics. 

For the implementation in a cold atomic setup, we consider previous experiments of ultracold rubidium atoms in optical lattices, where Fourier synthesized lattice potentials are used in order to tailor the atomic dispersion relation~\cite{salger2007atomic,salger2011klein}.
A trapping potential for atoms can be realised by superimposing a dipole potenial generated by a focused far red detuned laser beam. To minimize the interatomic interactions it will be desirable to operate with a moderate number of atoms, typically a few thousand.
After initialising, the momentum of the atoms must be manipulated, in order to produce relevant states of the simulated qubit and bosonic mode. 

Notice that the qubit state is encoded in the occupation of the Bloch bands $\ket{\pm 2\hbar k_0}=\ket{n_{\text{b}}\cdot4\hbar k_{0}-2\hbar k_{0}}\equiv\ket{n_{\text{b}}}$, while the bosonic mode quadratures are encoded in the position $\hat x$ and quasi-momentum $\hat q$ of the atoms. The qubit can be initialized in an arbitrary state by preparing the atoms in the corresponding position of the Bloch spectrum. This can be done by applying a Doppler-sensitive Bragg pulse~\cite{Grossert2015}.
Due to momentum conservation, the process entails a discrete momentum kick of $\pm 2\hbar k_0$. By controlling the share of atoms that gain  positive or negative momentum, as well as the relative phase between the Bragg pulses, the effective qubit can be initialized in any superposition of  $\sigma_x$ eigenstates [as defined in Eq.~\eqref{pauliformalism}].

Both the momentum (and correspondingly the state of $\sigma_{\text{x}}$), and in principle the position can be measured with absorption imaging techniques~\cite{Bloch2008}. For the former, standard time-of-fligth imaging can be used, as performed by simultaneously deactivating both the lattice beams and the dipole trapping potential and then detecting the atoms in the far field after a given free expansion time. While the reconstruction in this way is possible with a high precision~\cite{salger2011klein}, achieving the required spatial resolution for an in situ position detection of the oscillation is experimentally challenging.

For completeness, we mention that the qubit operator $\sigma_z$ can also be directly measured via adiabatic mapping~\cite{kastberg1995adiabatic, Bloch2008}, but only when the system state is close to the avoided crossing in the Bloch spectrum. By accelerating adiabatically the lattice from $q\sim 0$ to $q\sim 0.5\hbar k_0$,  Bloch waves are mapped onto the free-particle momentum states. Such a process corresponds to a rotation in the effective qubit Hilbert space. The required adiabatic acceleration of the mapping can be performed by means of a continuous frequency chirp applied to one of the lattice laser beams.

\begin{figure}[]
\centering
\includegraphics[angle=0, width= 0.5\textwidth]{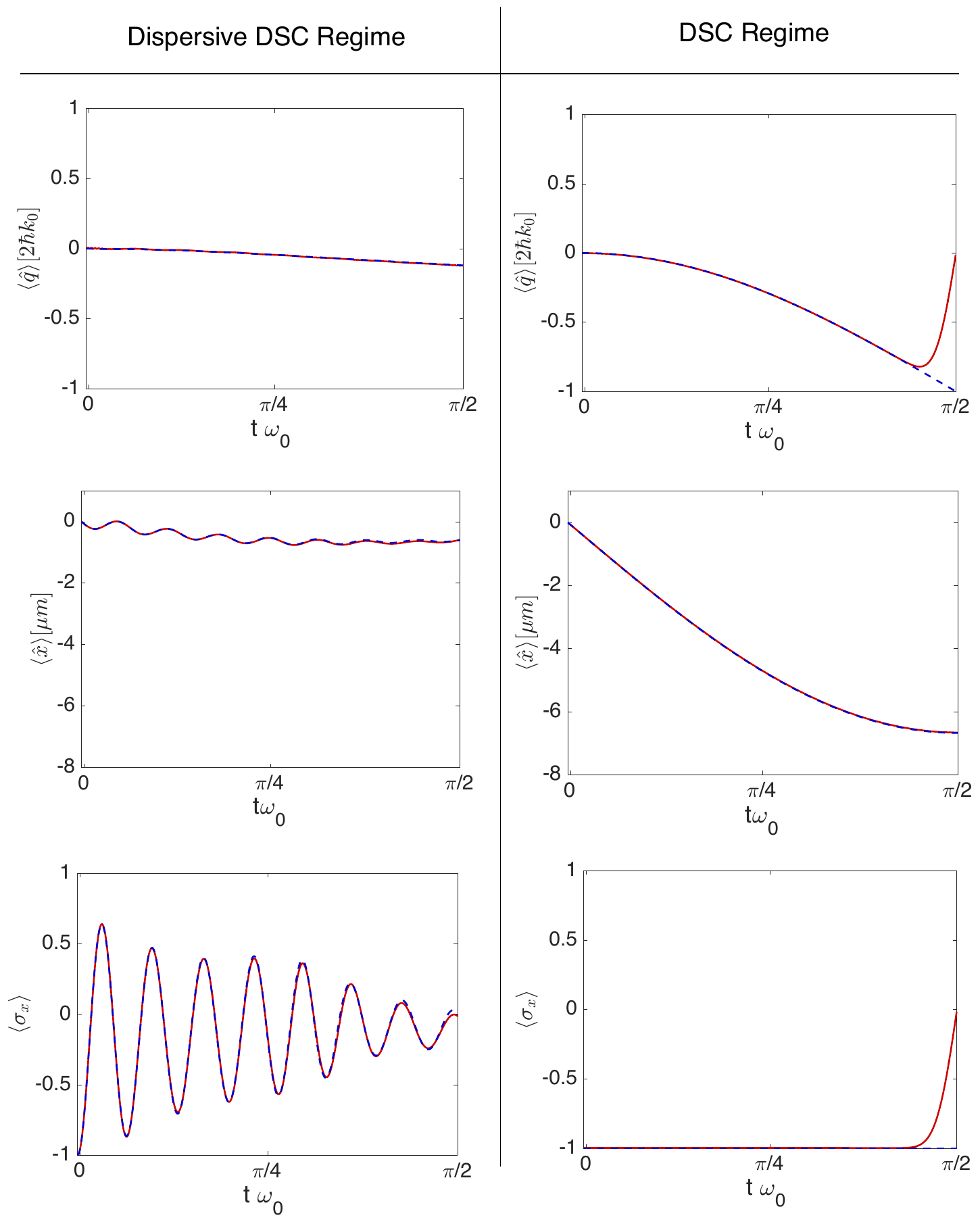}
\caption{\label{crossover}Comparison between the full cold atom Hamiltonian (red continuous line) and the quantum Rabi model (blue dashed line). The momentum is shown in units of $\hbar k_0$, while the position in units of $1/k_0$ and the coupling strength $g/\omega_{0}=5.18$. For the dDSC, the ratio of frequencies is given by $\omega_q/\omega_0 = 28.7$ while, for the DSC regime,  $\omega_q/\omega_0 = 0$.}
\end{figure}

Let us now consider the dynamics of the quantum Rabi model in the specific parameter regimes of interest for the proposed quantum simulation. Given that only very high values of the ratio between the coupling strength $g$ and the bosonic mode frequency $\omega_0$ are accessible, the rotating wave approximation can never be applied and the model cannot be implemented in  the Jaynes-Cummings limit. However, we will show that interesting dynamics at the crossover between the dDSC and the DSC regime can be observed, for values of parameters that are unattainable with so far available natural implementations of the quantum Rabi model.

By means of numerical simulations, we have compared the dynamics of the full cold-atoms model in Eq.~\eqref{fullHam} with the corresponding effective quantum Rabi model in Eq.~\eqref{standardRabi}. Numerical simulations of the full model have been performed in the position basis, applying a discretization of the real space over more than $10^3$ lattice sites. The quantum Rabi model has been numerically simulated introducing a cut-off ($N>500$) on the maximum number of bosonic excitations.

In Fig.~\ref{crossover}, we show the results of such numerical simulations, in different parameter regimes. The initial state $\ket{\psi_0} = \ket{q=0\hbar k}\ket{n_{\text{b}}=1}$ is given by the vacuum of the bosonic mode and an eigenvectors of $\sigma_x$. Such a state can be obtained preparing the atomic cloud at the center of the harmonic trap and at the $ q = 0$ of the $n_b= 1$ band of the Bloch spectrum, which corresponds to atoms prepared at $p=+2\hbar k$. In all plots, the red continuous line shows the dynamics of the full model (cf. Eq.~\eqref{fullHam}), while the dashed blue line corresponds to the quantum Rabi model (cf. Eq.~\eqref{standardRabi}). The good agreement between the two simulations breaks down when the system state hits the border of the validity region of the quantum simulation.  Different behaviors between the two regimes are more visible in the expected value of $\sigma_x$ which, in the DSC regime, is approximatively a conserved quantity, as shown below.

\begin{figure}[]
\centering
\includegraphics[angle=0, width= 0.38\textwidth]{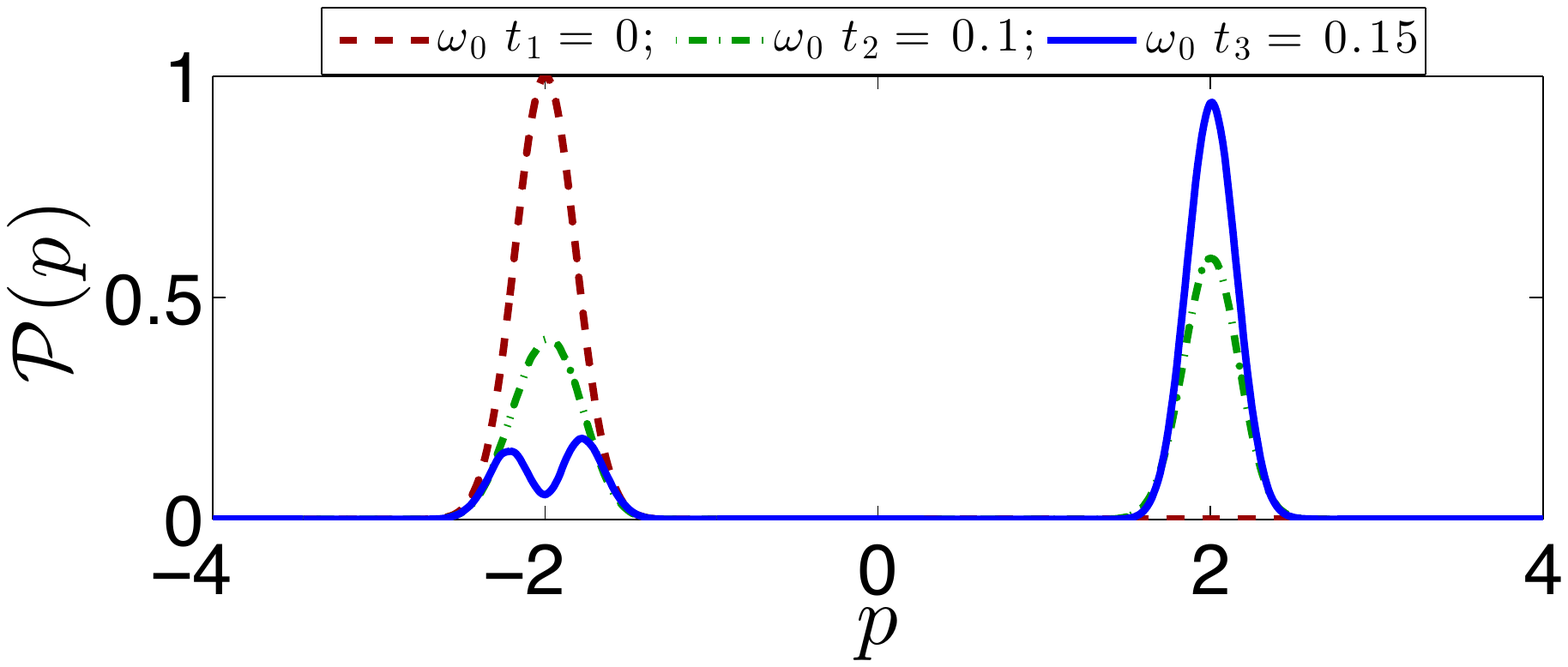}
\includegraphics[angle=0, width= 0.38\textwidth]{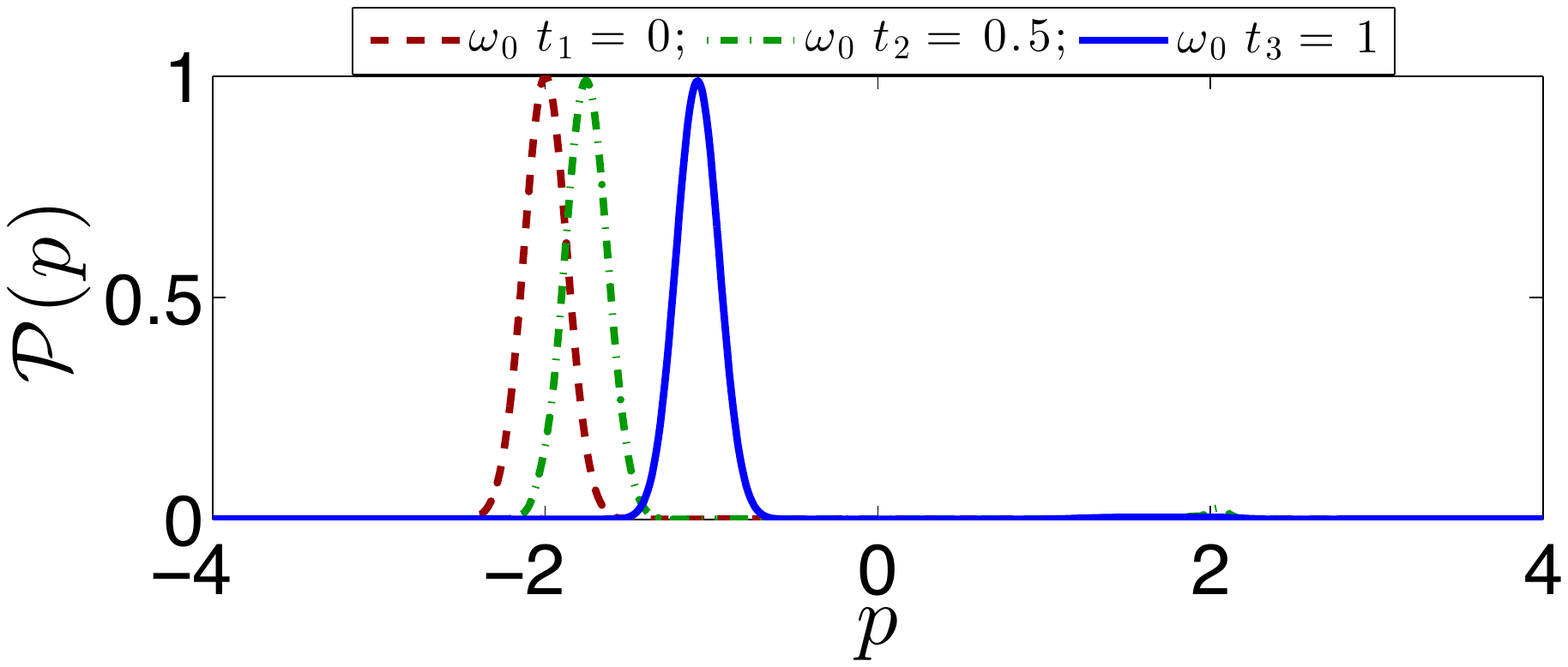}
\caption{\label{realmom}  Numerically evaluated real-momentum distribution of the cold atom cloud during the dynamics showed in Fig.~\ref{crossover}, at different evolution times.  For the dDSC regime (upper panel), the Rabi parameters are given by $g/\omega_0 = 7.7$ and $g/\omega_q = 0.43$. In this case, the initial wave-function is transformed back and forth between two distributions centered on the states $\ket{p=\pm2\hbar k_0}$. For the DSC regime (lower panel),  $g/\omega_0 = 10$ and $\omega_0 = \omega_q$. In this case, the system is continuously displaced in momentum space till a maximum value of the momentum.}
\end{figure}

In Fig.~\ref{realmom}, it is shown the  distribution $\mathcal{P}(p) = |\langle p \ket{\psi(t)}|^2$of the atomic physical momentum $\hat p$, for different evolution times. The momentum distribution can be experimentally obtained via time-of-flight measurements, and gives a clear picture of the system dynamics during the quantum simulation of the quantum Rabi model. The cloud is initialized in the momentum eigenstate $\ket{p=-2\hbar k_0} =\ket{q=0}\ket{n_{\text{b}}=0}$. When the periodic lattice strength $V$ is large enough, the dynamics is dominated by the coupling between the  $\ket{p=\pm2\hbar k_0}$ states. This case corresponds to the dDSC regime in Fig.~\ref{crossover}. Otherwise, the dynamics is dominated by the harmonic potential, and the evolution resembles the quantum Rabi model in the DSC regime.

\begin{figure}[]
\centering
\includegraphics[angle=0, width=0.47\textwidth]{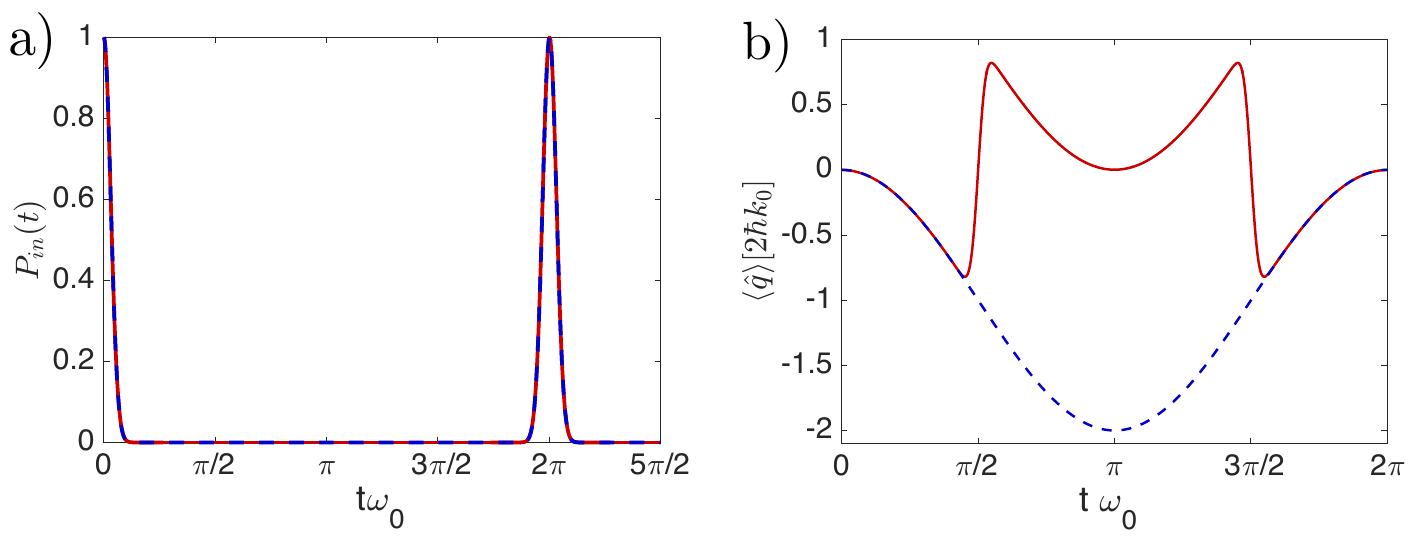}
\caption{\label{CandR}Comparison between the full cold atom Hamiltonian (red continuous line) and the QRM (blue dashed line). a) The plot shows collapses and revivals of the initial-state population $P_{in} = |Ê\langle \psi_{in}\ket{\psi(t)}|^2$. The initial state is given by  $\ket{\psi_{in}} = \ket{q=0}\ket{n_{\text{b}}=1}$. The coupling strength is given by $g/\omega_{0}=5.18$, while the qubit energy spacing vanishes $\omega_q = 0$. In this limit, collapses and revivals corresponds to harmonic oscillations of the atoms in the trap potential.
b) Temporal evolution of the quasi-momentum in units of $2 \hbar k_0$, as in right column of Fig.~\ref{crossover}, for the long time dynamics. Notice that  the value of the quasi-momentum is bound at $|q| \le 2 \hbar k_0$ and that the different behavior between the Rabi and the periodic quantum Rabi model appears at the boundary of the Brillouin zone.}
\end{figure}

We have first shown that our proposal is able to reproduce the dynamics of the quantum Rabi model at the crossover between the dDSC and DSC regimes. The analogy is broken when the value of the simulated momentum exceeds the borders of the first Brillouin zone. When this is the case, the model represents a generalization of the quantum Rabi model in a periodic phase space. In the following we show that collapses and revivals of the initial state, which represent the signature of DSC regime of the quantum Rabi model, are matched by the full atomic model.
Let us first review the quantum Rabi model dynamics, considering the initial state $\ket{\psi_{in}} = \ket{q=0}\ket{n_{\text{b}}=0,1}$. In the DSC regime, the system evolution is described by the approximated solution~\cite{Casanova2010},
\begin{equation}
\ket{\psi(t)} = e^{-i \hat H_{DSC} t  } \ket{\psi_{in}} = -1e^{-i\phi(t)} D\big[(-1)^{n_{\text{b}}}\beta(t)\big] \ket{0}\ket{n_{\text{b}}}
\label{eq_CandR}
\end{equation}
where $\phi(t)$ is a global phase independent of the qubit state, while $D[\beta(t)]$ is a displacement operator and $\beta(t) =  i\frac{g}{\omega_0}\left( e^{-i\omega_0 t} - 1\right) $. Accordingly, during the system time evolution $\sigma_z$ is conserved, while the vacuum state is displaced into a coherent state that rotates in phase space and that returns into the initial state with period $T= 2\pi/\omega_0$. This pattern of collapses and revivals is shown (blue dashed line) in Fig.~\ref{CandR}(a)  for the case in which the solution of Eq.~\eqref{eq_CandR} is exact ($\omega_q=0$). The width of the peaks is given by the width of the  momentum distribution of the initial state.  As shown in Fig.~\ref{CandR}(b) the periodicity of the momentum results in shifted values of $\hat q$ for half period of the system dynamics. Notice that, when the system is initialized in the state $\psi_{in} =  \left(  \ket{q=0}\ket{n_{\text{b}}=1}+\ket{q=0}\ket{n_{\text{b}}=0}\right)/\sqrt{2}$, the dynamics of Eq.~\eqref{eq_CandR} leads to Schroedinger cat generation  $\psi(t) =\left( \ket{\beta(t)} \ket{n_{\text{b}}=1} + \ket{-\beta(t)} \ket{n_{\text{b}}=0}\right)/\sqrt{2}$. The size of the cat state is given by the maximum value of the displacement $\beta_{\rm max} = 2\frac{g}{\omega_0}$, and so it is proportional to the coupling strength.

We have developed a method to implement a quantum simulation of the quantum Rabi model for unprecedented values of the coupling strength, using a system of cold atoms in a periodic lattice. Furthermore, the proposed scheme represents a generalization of the quantum Rabi model in the first Brillouin zone of the periodic phase space. A natural extension of the present work is the inclusion of atomic interactions, in order to implement a many-body system composed of interacting quantum Rabi models.

We acknowledge useful discussions with Michele Modugno. This work was supported in parts by the DFG (We 1748-20). SF acknowledges funding from University Sorbonne Paris Cit\'e EQDOL contract, while ER and ES from UPV/EHU UFI 11/55, MINECO FIS2015-69983- P, and UPV/EHU Project No. EHUA15/17. CS  acknowledges funding from Fundación General CSIC (Programa ComFuturo).

\end{document}